\documentclass[12pt]{article}
\pdfoutput=1
\usepackage{graphicx} % Required for inserting images

\usepackage{amssymb}
\usepackage{amsmath}
\usepackage[normalem]{ulem}
\usepackage{latexsym}
\usepackage[usenames]{color}
\usepackage{fancybox}
\usepackage{simplewick}
\usepackage{comment}
\usepackage{cite}
\usepackage{framed}
\definecolor{shadecolor}{rgb}{0.9,0.9,0.95}
\usepackage{setspace}
\usepackage{pdflscape}
\usepackage{multirow}
\usepackage{tabularx,booktabs}
\usepackage{bm}
\definecolor{darkgreen}{rgb}{0,0.5,0}
\definecolor{darkblue}{cmyk}{0.9,0.9,0,0}
\definecolor{darkred}{rgb}{0.6,0,0.3}

\usepackage{graphicx}
\definecolor{EdwardsLinkColor}{rgb}{0.0,0.0,0.3}

\usepackage[
  unicode=true,
  bookmarks=true,
  bookmarksnumbered=false,
  bookmarksopen=false,
  breaklinks=false,
  pdfborder={0 0 0},
  pdfborderstyle={},
  backref=false,
  colorlinks=true,
  linktocpage=true,
  linkcolor=EdwardsLinkColor,
  citecolor=EdwardsLinkColor,
  urlcolor=EdwardsLinkColor,
  filecolor=EdwardsLinkColor,
  anchorcolor=EdwardsLinkColor,
  menucolor=EdwardsLinkColor,
]{hyperref}

\hypersetup{pdftitle={OCO-Triality Beyond Matrix Models}}

\newcommand{\tr}{{\rm tr}}

\renewcommand{\thefootnote}{\arabic{footnote}}

\def\del{\partial}

\def\eqref#1{(\ref{#1})}
\def\comma{\,,}
\def\period{\,.}

\def\beq{\begin{equation}}
\def\eeq{\end{equation}}

\numberwithin{equation}{section}
\let\mc=\mathcal
\let\mb=\mathbb

\newcommand\be{\begin{equation}}
\newcommand\ee{\end{equation}}

\newcommand{\inv}{\frac{1}}
\newcommand{\half}{\frac{1}{2}}

\newcommand{\bea}[1]{\begin{eqnarray}\label{#1} }
\newcommand{\eea}{\end{eqnarray}}

\newcommand{\bdel}{\bar{\partial}}

\newcommand{\ga}{\gamma}
\newcommand{\de}{\delta}

\newcommand{\ds}[1]{\textcolor{blue}{[\textbf{DS}: #1]}}

\newcommand{\dg}{\dagger}

\renewcommand{\a}{\alpha}

\renewcommand{\b}{\beta}

\renewcommand{\l}{\lambda}
\newcommand{\s}{\sigma}

\newcommand{\psid}{\psi^{\dg}}

\newcommand{\cp}{\mb{CP}^{1}}
\newcommand{\bx}{\bar{x}}
\newcommand{\by}{\bar{y}}
\newcommand{\tX}{\tilde{X}}
\newcommand{\tZ}{\tilde{Z}}
\newcommand{\ty}{\tilde{y}}

\begin{document}

\title{\textbf{Open-Closed-Open Triality\\ Beyond Matrix Models}}

\author{Edward A. Mazenc${^{\beta}}$ \&  Debmalya Sarkar${^{\gamma}}$}
\date{}

\maketitle
\begin{comment}
  \makeatletter{\renewcommand*{\@makefnmark}{}
\makeatletter{\renewcommand*{\@makefnmark}{}
\footnotetext{${}^\beta$emazenc@ethz.ch}\makeatother}
\makeatletter{\renewcommand*{\@makefnmark}{}
\footnotetext{${}^{\gamma}$dsarkar@g.harvard.edu}\makeatother}  
\end{comment}

\begingroup
\renewcommand{\thefootnote}{}
\footnotetext{$^{\beta}$ emazenc@ethz.ch}
\footnotetext{$^{\gamma}$ dsarkar@g.harvard.edu}
\endgroup

\begin{center}
  \small{${}^\beta$\it{Institut für Theoretische Physik, ETH Zürich, \\ CH-8093 Zürich, Switzerland.}}\\
  \vspace{0.1cm}
  \small{${}^{\gamma}$\it{Jefferson Physical Laboratory, Harvard University, \\Cambridge, MA 02138, USA.}}
\end{center}

\vspace{2cm}

\begin{abstract}
\noindent 
We explore the ideas of open-closed-open triality within twisted holography. Starting from two transverse stacks of branes in the B-model on the resolved conifold, we obtain two equivalent open string descriptions. Both are full-fledged field theories. In the appropriate limit, they simplify to gauged $\beta\gamma$-systems with determinant insertions, as expected from open string field theory. The $\rho$-matrix models that have appeared previously in the literature appear from an on-shell analysis of the field-theory actions. Most importantly, we show that the effective potential generated by integrating out the open strings on the other stack of branes exactly captures their backreaction on the geometry. In particular, we match the action of the probe branes to that of giant gravitons in the \textit{deformed} conifold. The second open string description of the triality thus directly diagnoses the emergent bulk spacetime. 

\end{abstract}

\pagebreak

\tableofcontents

\pagebreak

\pagebreak

\section{Overview \& Main Results}
In this paper, we explore the ideas of open-closed-open triality \cite{Joburg,gopakumar2023deriving, Gopakumar:2024jfq} in the context of twisted holography \cite{CostelloGaiottoTH}. Twisted holography is an equivalence between 
\begin{itemize}
    \item a gauged $\beta\gamma$ system with fields transforming in the adjoint representation of $U(N)$ %(equivalent to a 2D chiral algebra $\mc{A}_N$ subsector of $\mathcal{N}=4$ Super Yang-Mills)
    \item a closed B-model topological string theory on the deformed conifold, $SL(2,\mathbb{C}) \simeq AdS_3 \times S^{3}$.
\end{itemize}
This gauge theory represents the full B-model open string field theory (OSFT) of $N$ topological $D1$ branes wrapping the blown up $\mb{CP}^1$ of the resolved conifold\footnote{Note that the original construction of \cite{CostelloGaiottoTH} starts with branes in $\mb{C}^3$, not in the resolved conifold. As recently discussed in \cite{Jarov:2025qhz}, one may alternatively start from the resolved conifold. The two fields parametrizing transverse fluctuations of the branes in the $\mathcal{O}(-1)$ fiber directions then transform as symplectic bosons. }. Unlike the full AdS/CFT correspondence, no additional low-energy or near-horizon limit of the brane dynamics is required in this topological setting. The $N$ $D1$-branes backreact on the geometry, giving rise to the deformed conifold as the emergent closed string bulk. This duality can be viewed in the same spirit as the A-model duality of \cite{GV99, OV2002worldsheet} and subsequent B-model constructions \cite{DijkgraafVafa02, DVtoprecKS, Aganagic_2005 }. The original construction of \cite{CostelloGaiottoTH} has inspired much work on other holographic descriptions of chiral gauge theories, including applications to flat-space and celestial holography \cite{topdownflat,Burns,Costello:2020jbh,Sharma:2025ntb, Costello:2022wso}. 

The gauged $\beta\gamma$-system furnishes a Lagrangian description of the two-dimensional $\mc{A}_N$ chiral algebra subsector of 4D $\mc{N}=4$ Super Yang-Mills theory with gauge group $U(N)$. As shown by \cite{beemRastelli2015}, it captures certain supersymmetry protected correlators in the physical theory. Twisted holography may therefore be also seen as a subsector of the well-known AdS/CFT correspondence, relating $\mc{N}=4$ SYM and type IIB string theory in $AdS_5 \times S^5$ \cite{Maldacena:1997re}.

The basic premise of open-closed-open triality is to consider two sets of branes and relate their open string descriptions. The most ambitious version of open-closed-open triality is the claim there may exist two fully equivalent open string descriptions of the same closed string background. The $(2,1)$ minimal string (also known as pure topological gravity \cite{Witten:1990hr}) provides such an example. On one hand, it is dual to the double-scaling limit of the Gaussian matrix model. The eigenvalues of the matrix are interpreted as the position of D-branes which give rise to this closed string background. Moreover, single trace operators map on to closed string vertex operators. On the other hand, the second open string description is given by the Kontsevich matrix model \cite{KontsevichAiry}, which was shown by Gaiotto and Rastelli to be the full open string field theory on FZZT branes, in the $(2,1)$-minimal string \cite{Gaiotto:2003yb}. These FZZT branes are treated as probes in the closed string geometry. It serves as a generating function of the closed string observables, roughly by re-expressing the perturbative backreaction of the FZZT branes as exponentials of (sums of) closed string vertex operators. 

The authors of \cite{exact} later showed the equivalence of these two matrix model descriptions by considering correlators of determinants in the Gaussian matrix integral and integrating in and out various degrees of freedom. What was later shown in \cite{Joburg, komatsuOCO,gopakumar2023deriving, Gopakumar:2024jfq} was that these manipulations implemented a dynamical graph duality of the Feynman diagrams of each theory, exchanging vertices and faces. Since single trace operators in the Gaussian corresponds to external vertices from a diagrammatic perspective, and those map onto the insertion of dual closed string vertex operators, it was dubbed as a V(ertex)-type duality. Instead, it is the faces of the Kontsevich model ribbon graphs that correspond to marked points on the worldsheet, earning its name as a F(ace)-type duality. The interplay between open-closed-open triality and a worldsheet dual for $\mathcal{N}=4$ SYM has also been discussed recently in \cite{Berkovits:2025xok, Berkovits:2026ifu}.

We would like to mimic the above scenario in the context of twisted holography. We will not realize the strong version of open-closed-open triality in this paper, in the sense that the second description does not serve as a generating function of all closed string observables ,as in the Kontsevich model. Our more humble starting point is to insert $Q$ $D1'$ branes, transverse to the $N$ original branes wrapping the $\mb{CP}^1$ in the resolved conifold. From the point of the $N \times N$ gauged $\beta \gamma$ system, these represent $Q$ determinant operator insertions. This is the analog of inserting determinants in the Gaussian matrix model. These $N$ branes backreact on the resolved conifold geometry to give rise to the deformed conifold $SL(2,\mathbb{C})$. The $Q$ $D1'$ will instead play the role of the FZZT branes in the $(2,1)$-minimal string example above.\footnote{There is an interesting similarity to the dual description on compact versus non-compact branes in the B-model dualities of \cite{Aganagic_2005}.} Starting from this system of two branes, we then derive three main results. 

\begin{figure}
    \centering
    \includegraphics[width=0.7\linewidth]{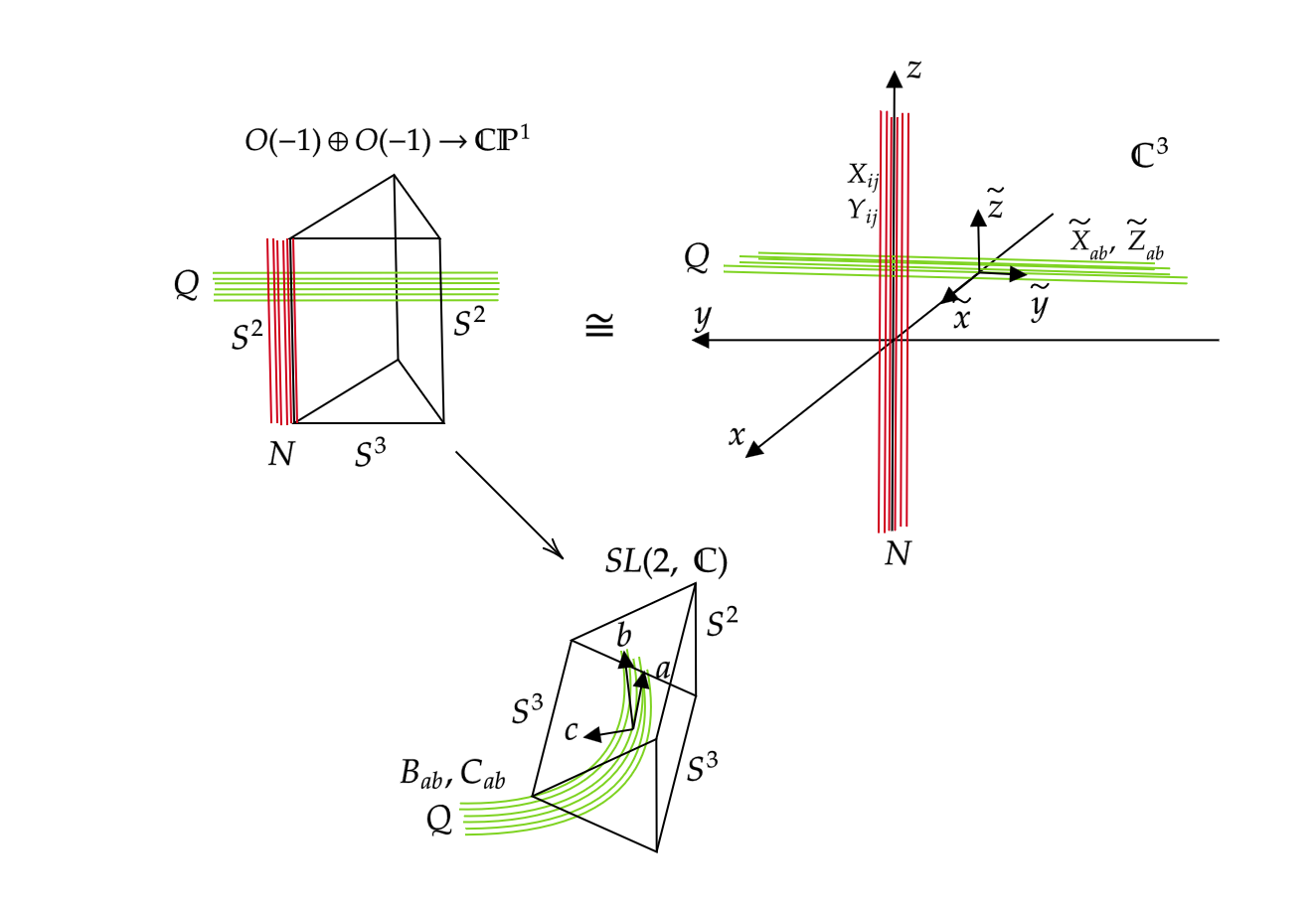}
    \caption{Summary of main results 1 and 2. (Left) A stack of $N$ branes (red) and a stack of $Q$ branes in the resolved conifold. (Right) Same setup in the $\mb{C}^3$ coordinates. We find two equivalent \textit{field theoretic} descriptions of the system, in terms of $N\times N$ fields $X, Y$ or $Q \times Q$ fields $\tX, \tZ$. (Bottom) We can also recover the backreacted picture, where we have $Q$ giant graviton branes in the deformed conifold with $Q\times Q$ fields $B, C$ capturing their fluctuations. }
    \label{fig:res12}
\end{figure}

\textbf{Main Result 1:} We show that this system is described by two independent \textit{field theories}, each corresponding to the dynamics of open strings extended between one set of branes, after integrating out the open strings on the other set. One of the main motivations of this short paper is to provide a counter-example to the observation that previous instantiations of open-closed-open triality seem to lead to dual matrix model descriptions. The two descriptions are the following
\be \label{eq:maineq1a}
\begin{split}
    &\inv{Z_N} \int DXDY\, \exp\bigg[ -N\int d^2z\, \tr_N(X\bdel Y) \bigg]\prod_{a=1}^Q \det(m_a + X(z_a) + u_aY(z_a)) \\
    &= \inv{Z_Q} \int D\tX D\tZ \exp\bigg[-N\int d^2\ty\, \tr_Q(\tX\bdel\tZ)\bigg] \det(m_a\de_{ab} - \tX_{ab}(\ty =z_b) + u_b\tZ_{ab}(\ty=z_b))^N,
\end{split}
\ee 
which is better understood in the coincident limit of the probe $D1'$ branes 

\be \label{eq:maineq1b}
\begin{split}
    &{\inv{Z_N} \int DXDY\, \exp\bigg[ -N\int d^2z\, \tr_N(X\bdel Y) \bigg] det(m + X(z_0))^Q}\\
    &\hspace{50pt}={\inv{Z_Q} \int D\tX D\tZ \exp\bigg[-N\int d^2\ty\, \tr_Q(\tX\bdel_{\ty}\tZ)\bigg] det(m - {\tX}(\ty=z_0))^N} .
\end{split}
\ee 
In the first line of \eqref{eq:maineq1b}, $X, Y$ are $N\times N$ matrix valued fields capturing the transverse fluctuations of open strings on $N$ branes, while the $Q$ determinants represent the $Q$ coincident probe branes. In the second line, the $Q\times Q$ matrix valued fields $\tX, \tZ$ describe the open strings between the probe branes, while the open strings on the $N$ original branes have been integrated out, giving rise to the $N$ determinant insertions. $Z_N$ and $Z_Q$ are normalization factors, corresponding to the path integral without determinant insertions. For the non-coincident limit \eqref{eq:maineq1a}, the geometric interpretation of the RHS is not very clear to us. In the coincident limit \eqref{eq:maineq1b}, see Fig. \ref{fig:res12} for the geometric picture. 

What is physically important about each field-theory description in \eqref{eq:maineq1b} is they descend directly from the full open string field theory of the branes in the original resolved conifold geometry, generalizing the previous matrix model descriptions.

\textbf{Main Result 2:}
We show that the field theory on the $Q$ probe branes is exactly equal to a $\beta\gamma$ system in the backreacted, deformed conifold geometry. In other words, the effective potential generated by integrating out the $N$ original branes can be recast geometrically. The authors of \cite{Budzik:2021fyh} discussed this setup in the context of an effective matrix model used to express the probe brane correlator. Moreover, they showed how solutions to the matrix model equations of motion could be mapped to the shape of the branes in the $SL(2,\mb{C})$ bulk. Here, we directly rewrite the action for the probe branes, derived from open-closed-open triality, as the action of giant gravitons in $SL(2,\mathbb{C}) \sim AdS_{3} \times S^{3}$. For the simplest geometric setup, this reduces to the equality of two actions (see Fig. \ref{fig:res12}):

\be 
\begin{split}\label{eq:maineq2}
    N\underbrace{\int d^{2}\ty \ \mathrm{tr}_Q \left( \tX \partial_{\bar{y}} \tZ\right) - N \mathrm{tr}_Q \log(\tX(0))}_{\text{Effective Action in Original Geometry}} & =   N\int \frac{d^2\ty}{\ty}\tr_Q\Big(\tX(\ty)\bdel_{\ty}\Big(\ty\tZ(\ty) -\inv{\tX(\ty)}\Big)\Big) \\
   &\hspace{30pt}=   \underbrace{\frac{N}{\pi}\int \frac{d^{2}a}{a} \ \mathrm{tr}_Q\left( B \partial_{\bar{a}} C\right),}_{\text{Action in Backreacted Geometry}}
\end{split}
\ee 

where $B_{ab},C_{ab}$ are also $Q\times Q$ fields, capturing the transverse fluctuation of the giant graviton branes in the backreacted geometry. This rewrite follows from a study of the coordinates in the original and backreacted geometries. For the resolved conifold $\mathcal{O}(-1) \oplus\mathcal{O}(-1)  \rightarrow \mathbb{CP}^1 $, we let $z$ parametrize the $\mathbb{CP}^1$ and $x$ \& $y$ the $\mathcal{O}(-1)$ fibers. In this simple geometric setup, the $Q$ probe branes are extended along the $y$ direction, at $x=z=0$. The original $N$ branes lie at $x=y=0$, giving rise to the source term $\mathrm{tr} \log(\tilde{X}(0))$.  
We can define new coordinates via the combination
\begin{equation}
    a \leftrightarrow  y \quad b \leftrightarrow  x \quad c \leftrightarrow  yz-\frac{1}{x} \quad d \leftrightarrow  xz
\end{equation}

and see they satisfy $ad-bc=1 $. These are thus coordinates on the backreacted geometry, i.e. the deformed conifold. It had already been noted by \cite{SuperpotforCS,DijkgraafVafa02} that a potential term for a gauged $\beta \gamma$ system could encode a change in complex structure, as also recently highlighted by \cite{Jarov:2025qhz}. Our work shows the effective potential, $\mathrm{tr} \log(\tilde{X}(0))$, generated by integrating out the open strings on the original $N$ branes indeed precisely reproduces their backreaction on the geometry. It is a purely open string diagnostic of the geometric transition from the resolved to the deformed conifold, the closed string theory target of twisted holography. The authors of \cite{Sharma:2025ntb} have also recently shown this same effective potential could be described in terms of an open-closed coupling between the open string field theory on the probe branes and the closed string Kodaira-Spencer field. 

%The field theory captures both the classical shape predicted by \cite{Budzik:2021fyh} and the transverse fluctuation of the open strings around that classical saddle. %In this way, we are able to understand the backreaction of the $N$ original branes on the resolved conifold using open strings/branes as probes. This indirectly represents the closed string picture in open-closed-open triality of \cite{gopakumar2023deriving}. 

\textbf{Main Result 3:} Ideas of open-closed-open triality were first used in \cite{komatsuOCO} to compute correlators of $Q$ half-BPS determinant operators in free $\mathcal{N}=4$ Super Yang-Mills (SYM), which correspond to insertions of giant graviton branes in the bulk. They derived an effective $Q \times Q$ matrix model, known in the literature as the $\rho$-matrix integral. This technique has found many subsequent applications, both in twisted holography \cite{Budzik:2021fyh} and self-dual $\mathcal{N}=4$ SYM \cite{Caron-Huot:2023wdh, Sharma:2025ntb}. While it was clear from the $Q \times Q$ index structure that the matrix $\rho$ described the open strings on the probe branes, it was not clear (at least to us) how this description was related to more standard DBI- or OSFT-actions. The third result of this paper is to show how the $\rho$-matrix model in twisted holography emerges from the on-shell action of our field-theoretic description: 
\be \label{eq:maineq3a}
\begin{split}
    \underbrace{N\int d^2\ty\, \tr_Q(\tX\bdel_{\ty} \tZ) - N\tr_Q\log(m - \tX(z_b) + u_b\tZ(z_b))\vert_{on-shell}}_{\text{on-shell action of the F-type field theory}} = \underbrace{-\frac{N\pi}{2}\frac{ z_{ab}}{u_{ab}}\rho_{ab}\rho_{ba} - N\tr_Q \log(m_a\de_{ab} + \rho_{ab})}_{\text{F-type matrix model}}.
\end{split}
\ee 
Using the equations of motion for the two fields $\tX$ and $\tZ$, we can relate them on-shell to the $\rho$-matrix:
\be \label{eq:maineq3b}
\tX_{ab}(\ty)\vert_{on-shell}= u_a\tZ_{ab}(\ty)\vert_{on-shell} = \frac{u_a}{\ty-z_a}\frac{z_{a}-z_{b}}{u_{a}-u_{b}}\rho_{ab}.
\ee 
Since the two fields $\tX$ and $\tZ$ are related to each other on-shell, we end up with an effective matrix model with only one matrix. Following the work of \cite{Budzik:2021fyh}, we see that the saddles of our F-type field theory determine the shape of giant graviton branes in the deformed geometry, and the higher order terms in the action characterize their fluctuations. 

The paper is structured as follows. In section \ref{bg}, we give a short summary of some necessary background material, reviewing the 2D chiral algebra and its relation to open string (field) theory on the conifold. In section \ref{oco_inout}, we discuss the correlation function of determinant operators in the chiral algebra and its dual description in terms of branes in the resolved conifold. We show an equivalent description in terms of a different field theory by integrating in and out certain degrees of freedom. In section \ref{coinc} we show that the dual action represents the open strings fluctuating on the probe branes. We can further take into account the backreaction from the original branes and recover the expected $\b\ga$ system in the deformed geometry. In section \ref{matrix} we spell out the details of how this dual field theory reduces to the $\rho$-matrix model discussed previously in the literature. We mention how the original path integral over the $N \times N$ fields with determinant insertions may also be reduced to a matrix model, further emphasizing that the matrix vs. field theory nature of the dual descriptions is logically independent from the ideas of open-closed-open triality. In section \ref{conc} we summarize the results and discuss some future directions. In appendix \ref{sym} we revisit the case of free SYM and speculate some relation to giant graviton branes \cite{McGreevy:2000cw, Hashimoto:2000zp, Balasubramanian:2001nh} in $AdS_5\times S^5$ \footnote{S. Murthy has suggested to us how the on-shell DBI action on giant graviton branes in $AdS_5\times S^5$ might be matched against the dual matrix model actions of \cite{Caron-Huot:2023wdh,Gopakumar:2024jfq}.}. 

\section{Background on Twisted Holography}\label{bg}
\subsection{2d Chiral Algebra as a Subsector of $\mathcal{N}=4$ SYM}
The original derivation of the 2D chiral algebra as a subsector of 4D supersymmetric theories was explained in \cite{beemRastelli2015}. In a nutshell, starting with a four-dimensional $\mc{N}=2$ SCFT, we can find a nilpotent charge $\mb{Q}$, and a subgroup 
\be 
{sl}(2)\times \tilde{{sl(2)}} \subset {sl(4|2)}
\ee 
of the group of symmetries, such that the generators of ${sl(2)}$ are $\mb{Q}$ closed, and the generators of $\tilde{{sl(2)}}$ are $\mb{Q}$ exact. Furthermore the two ${sl(2)}$ act as the holomorphic and antiholomorphic Mobius transformation of a $\mb{C}\subset \mb{R}^4$. Fields in the nontrivial cohomology of $\mb{Q}$ can be represented as chiral fields in this $\mb{C}$, and their operator products define a 2D chiral algebra. 

\begin{figure}
    \centering
    \includegraphics[width=0.9\linewidth]{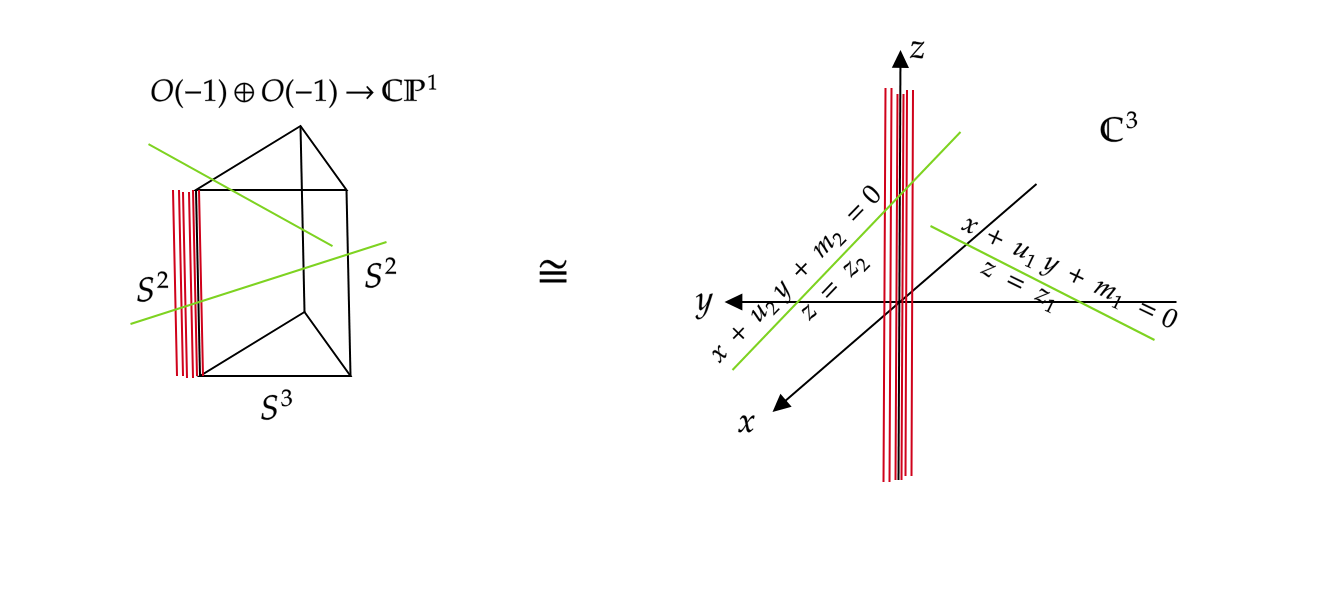}
    \caption{Topological branes in the resolved conifold. (Left) A stack of $N$ branes (in red) wrapping the blown up $\cp$, with a few probe branes (in green) that are transverse to the $N$ branes. (Right) In a patch of the resolved conifold described by $x, y, z \in \mb{C}$, the $N$ branes wrap the complex curve $x=0,y=0$, and the probe branes wrap the curve $x + u_a y + m_a =0, z=z_a$.}
    \label{fig:probe1}
\end{figure}

The chiral algebra $\mc{A}_N$ is obtained from passing 4D  $\mc{N}=4$ SYM with gauge group $U(N)$ through the above machinery. It is described by a theory of gauged symplectic bosons $X, Y$ and fermions $b, c$ transforming in the adjoint of $U(N)$. The gauge-fixed action is given by 
\be \label{gauged_bg}
S = N\int d^2z\ \tr_N(X\bdel Y+b\bdel c),
\ee 
with BRST charge 
\be 
Q_{BRST} \sim N\oint dz\ \tr_N(c[X,Y] +\half b[c,c] )
\ee 
and the following OPEs 
\be 
X_{ij}(z)Y_{kl}(0)\sim \inv{Nz}(\de_{il}\de_{jk}) \ \ ; \ \ b_{ij}(z)c_{kl}(0)\sim \inv{Nz}(\de_{il}\de_{jk}) 
\ee 
We can build a bosonic family of gauge invariant operators using the single letter 
\be 
\mc{W}(u,z) = X(z) + uY(z)
\ee 
satisfying the OPE 
\be 
\mc{W}_{ij}(u_a, z_a)\mc{W}_{kl}(u_b, z_b) \sim \inv{N}\frac{u_{ab}}{z_{ab}}(\de_{il}\de_{jk}).
\ee 
The parameter $u$ captures the transformation property of a field under the $SL(2,\mb{R})$ global symmetry algebra. At large $N$, the BRST closed operators are spanned by the single traces
\be \label{a_n}
A_n(u, z) = \tr_N(\mc{W}(u,z)^n)
\ee 
and their superconformal partners (involving the ghost fields), and their products. For our purposes we will restrict ourselves to the operators of the form \eqref{a_n}. To be precise, the main object of our interest will be (sub)determinant insertions 
\be 
det(m + \mc{W}(u,z))
\ee 
They correspond to the insertion of additional probe branes, see fig. \ref{fig:probe1}.

\subsection{Conifolds and Open String Field Theory}
The conifold is a 3 complex-dimensional manifold, described by a quadric in $\mb{C}^4$
\be \label{coni_1}
\sum_{i=1}^4 w_i^2 = 0 \Longrightarrow ad-bc=0.
\ee 
The point $\{w_i = 0\}$, or $a, b, c, d=0$ is a singular point. There are two different ways to de-singularize this point, known as resolution or deformation. 
\begin{enumerate}
    \item \textbf{Resolved conifold:} In this approach we replace the equation \eqref{coni_1} by the equation 
    \begin{gather}\label{res_con_1}
        \begin{bmatrix}
            a & b\\
            c & d
        \end{bmatrix}
        \begin{bmatrix}
            \l_1\\
            \l_2
        \end{bmatrix} = 0
    \end{gather}
    where $[\l_1, \l_2]\in \mb{CP}^1$, so we have defined a 3 complex dimensional manifold in $\mb{C}^4 \times \mb{CP}^1$. When we are away from the singular point, $\l_i$ are exactly determined in terms of $a,b,c,d$. But at the singular point, any $\l_i$ are allowed, so we have effectively blown up the singularity into a $\cp$. 

    Alternatively, we can define the following coordinates 
    \be \label{res_con_2}
    a = y , \ b = x ,\ c = yz , \ d = xz
    \ee 
    which satisfy the conifold condition, and as $x, y\to 0$, we see that $z$ is unconstrained, hence $\{x=0,y=0,z\}$  represents the blown up $\cp$. This coordinate system is linked to the other representation of the resolved conifold as total manifold of
    \be \label{res_con_3}
    \mc{O}(-1)\oplus \mc{O}(-1) \to \cp
    \ee 
    where $z$ can be thought of as the coordinate on the base manifold, and $x,y$ as the coordinates on the fibres. We mention in passing that in the $(x,y,z)$ coordinate system, the holomorphic 3-form of this Calabi Yau manifold takes a simple form 
    \be \label{res_con_omega}
    \Omega = dz\wedge dx\wedge dy
    \ee 
    which will be important for us later.    
    \item \textbf{Deformed conifold:} Another way to repair is to deform the equation \eqref{coni_1} to 
    \be \label{def_con}
    ad-bc = \mu,
    \ee 
    where $\mu$ is a constant. With this modification, the singular point is replaced by an $S^3$ of radius $\sqrt{|\mu|}$ instead of the $\mathbb{CP}^1 \sim S^2$ of the resolved conifold. Being a Calabi Yau 3-fold, the deformed conifold has a nowhere vanishing holomorphic 3-form $\Omega$, and the constant $\mu$ is proportional to the period of $\Omega$ along the compact $S^3$. On a patch $a\neq 0$, the holomorphic form can be written as 
    \be \label{def_con_omega}
    \Omega = \frac{da \wedge db \wedge dc}{a}
    \ee 
\end{enumerate}
\subsubsection*{Open String Field Theory on Conifolds} \label{sec:OSFT}
Calabi Yau 3-folds are a natural choice of target space for discussing B-model topological string theory. We are mainly interested in the target space description of open strings in this model. As shown by \cite{WittenCSasOSFT}, the open string field theory (OSFT) of $N$ space filling $D$ branes in CY3 $\mc{M}$ is described by the holomorphic Chern-Simons theory
\be\label{holo_CS}
S = \int_\mc{M} \Omega\wedge \tr(A\wedge \bdel A + \frac{2}{3}A\wedge A\wedge A)
\ee 
where $A\in \Omega^{0,1}(\mc{M})$. The situation we are interested in is the theory on lower dimensional $D$ branes, which can be obtained from \eqref{holo_CS} by dimensional reduction. For example, if we consider $D1$ branes wrapping a complex curve $\Sigma$ holomorphically embedded in $\mc{M}$, then the action of the system can be obtained from \eqref{holo_CS} and shown to be 
\be \label{holo_CS_bg}
S = \int_\Sigma d^2z\  \Omega_{zij} \tr(\b^i D_{\bar{z}}\b^j )
\ee 
where $D_{\bar{z}} = \bdel + [A_{\bar{z}}, \cdot ]$, $\beta^{i} (i= 1,2)\in N^{0,1}\Sigma$ represent fluctuations of the branes in the transverse direction, and $A \in T^{*(0,1)}\Sigma$. By gauge-fixing (to, say $A_{\bar{z}} = 0$) and including the ghost action, we see that the theory becomes identical to \eqref{gauged_bg}, i.e. a gauged $\b\gamma$ system. This is analogous to $\mc{N}=4$ SYM being the theory of open strings on D-branes wrapping $\mb{R}^{1,3}\subset \mb{R}^{1,9}$ \cite{Maldacena:1997re}. We will analyse the action \eqref{holo_CS_bg} for the resolved and deformed conifold in more details in the following sections.

\section{Probe Branes in Twisted Holography and Integrating in-out}\label{oco_inout}
We start with the resolved conifold
\be 
\label{res_con_3}
\mc{O}(-1)\oplus \mc{O}(-1) \to \cp
\ee 
with a stack of $N$ topological branes wrapping the $\mathbb{CP}^1$ along $x=y=0$, see Fig. \ref{fig:probe1}. Following the discussion in the previous section, we can write the action as 
\be 
    S = N\int d^2z \tr_N(X\bdel Y+b\bdel c),
\ee 
where the base is paramterized by $z\in \mb{CP}^1$, and $X(z), Y(z)$ are $U(N)$ adjoint fields that represent the fluctuation of the branes in the transverse fibre directions. Moreover we have used the holomorphic 3-form \eqref{res_con_omega}, and gauge-fixed to $A_{\bar{z}}=0$, hence including the ghost action. 

\begin{figure}
    \centering
    \includegraphics[width=0.7\linewidth]{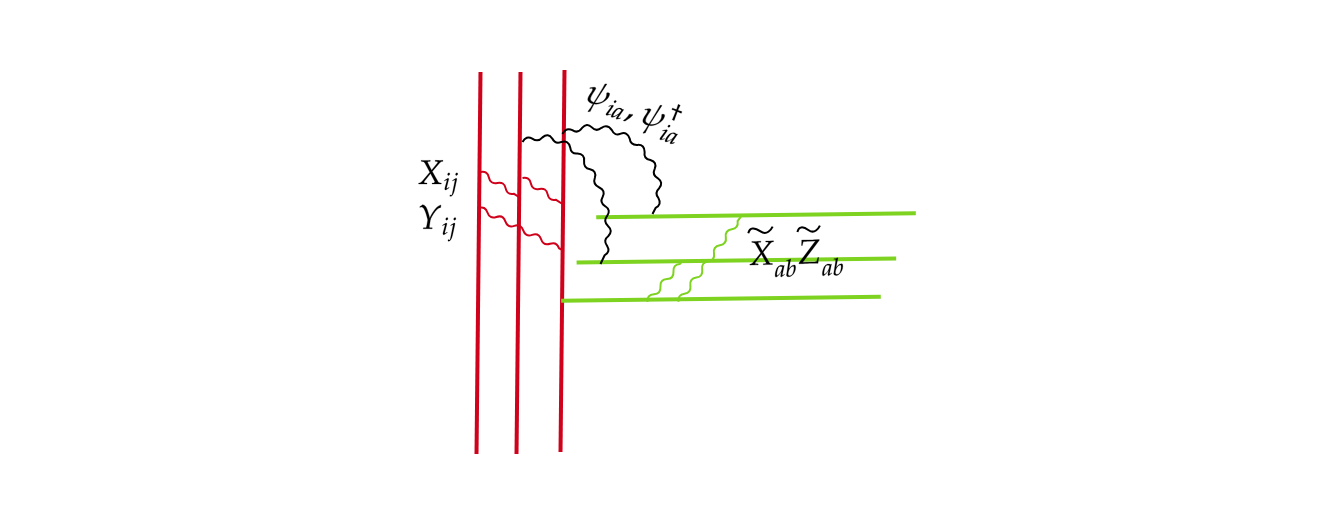}
    \caption{The two sets of branes in question and the different kinds of open strings stretched between them.}
    \label{fig:openstrings}
\end{figure}

We will consider the BRST closed (sub)determinant operators 
\be \label{subdet}
D(m, u, z) = \det(m + \mc{W}(u,z)) = \det(m + X(z) + uY(z))
\ee 
and look at their correlation functions 
\be \label{subdet_corr}
G_Q = \langle D(m_1, u_1, z_1) \dots D(m_Q, u_Q, z_Q)\rangle.
\ee 
Each determinant operator corresponds to inserting a $D1'$  brane along the curve $x + u_ay + m_a = 0, z= z_a$ in the resolved conifold, see Fig. \ref{fig:probe1}. Since we are considering  BRST closed operators involving no ghost-fields, we will ignore the ghost part of the action. \eqref{subdet_corr} is computed by the following path integral expression
\be \label{ch_Z}
G_Q = \inv{Z_N} \int DXDY\, \exp\bigg[ -\frac{N}{2}\int d^2z\, \tr_N(X\bdel Y - Y\bdel X) \bigg]\prod_{a=1}^Q \det(m_a + \mc{W}(u_a, z_a)),
\ee 
where we are using the symmetrized kinetic action. The objective of this section is to generalize the integrating in-out procedure of \cite{komatsuOCO, Budzik:2021fyh, gopakumar2023deriving, Gopakumar:2024jfq} and re-express \eqref{ch_Z} as a field theory on the probe branes. Let's go through the steps:

\paragraph{Integrating in fermionic bi-fundamentals:} We  can write the determinants in terms of fermionic variables in the bi-fundamental and anti-bi-fundamental representations of $U(N) \times U(Q)$\footnote{Using the relation $ \det(X) = \int d\psid_i d\psi_j \exp(\psid_i X_{ij}\psi_j)$}, leaving us with  
\be \label{ch_Z_psi}
G_Q = \inv{Z_N} \int DXDY d\psid_a d\psi_a \, \exp \bigg[ - \frac{N}{2}\int d^2z\, \tr_N(X\bdel Y - Y\bdel X) + \psid_{ia}(m_a \de_{ij} + X_{ij}(z_a)+ u_aY_{ij}(z_a))\psi_{ja} \bigg]\\
\ee 
Looking at the index structure, we can interpret the fermionic fields $\{\psi_{ia}, \psid_{jb}\}$ as open strings between the original stack of $D1$ branes and the probe $D1'$ branes, see Fig. \ref{fig:openstrings}. 
    
\paragraph{Integrating out the $N \times N$ adjoints:} Next we notice that the action is quadratic in $X \ \& \ Y$, in particular we can write it as 
\be 
\begin{split}
    G_Q &= \inv{Z_N} \int DXDY d\psid_a d\psi_a \exp\bigg[ -\frac{N}{2}\int d^2z\,\tr_N\Big[(X+ X_{0})\bdel (Y - Y_{0}) - (Y-Y_0)\bdel(X+X_0)\Big]\\
    &\hspace{170pt} -\frac{N}{2}\int d^2z\, \tr(X_0\bdel Y_0 - Y_0\bdel X_0) + \sum_a m_a\psid_{ak}\psi_{ak}   \bigg],
\end{split}
\ee 
where $X_{0, ij} = \inv{N}\sum_a \frac{\de(z-z_a)}{\bdel} u_a\psid_{aj}\psi_{ai}$, $Y_{0, ji} = \inv{N}\sum_b \frac{\de(z-z_b)}{\bdel}\psid_{bi}\psi_{bj}$\footnote{We have used the following shorthand: $\bdel\Big(\inv{z}\Big) = \pi \de(z) \ \implies \frac{\de(z)}{\bdel} := \inv{\pi z} $}, therefore 
\be 
\frac{N}{2}\int d^2z\, \tr(X_0\bdel Y_0 -Y_0\bdel X_0) = \inv{2N}\int d^2z\, \sum_{a\neq b} \frac{\de(z-z_b)}{\pi(z-z_a)}u_{ab}\psid_{aj}\psi_{ai}\psid_{bi}\psi_{bj} = \frac{1}{2N\pi}\sum_{a\neq b}\frac{u_{ab}}{z_{ab}}(\psid_{ia}\psi_{ib})(\psid_{jb}\psi_{ja})\footnote{\text{The diagonal terms ($a=b$) are divergent but they exactly cancel between the two terms.}}.
\ee 
So we can integrate out the adjoint fields $X$ and $Y$ and write the correlation function only in terms of the fermions 
\be \label{ch_psi}
G_Q = \int d\psid_a d\psi_a \exp\bigg[ - \inv{2N\pi}\sum_{a\neq b}\frac{u_{ab}}{z_{ab}}(\psid_{ia}\psi_{ib})(\psid_{jb}\psi_{ja}) + \sum_a m_a\psid_{ka}\psi_{ka} \bigg]
\ee 
This is the picture in terms of only the open strings extended between the original and probe branes.

\paragraph{Integrating in dual adjoint fields:} We want to express this quantity in terms of a field theory defined on the probe branes, i.e. in terms of only the open string modes on the $D1'$ branes. To do so, we integrate in the following quantity in \eqref{ch_psi} 
\be 
1= \inv{Z_Q}\int D\tX D\tZ \exp\Bigg[-\frac{N}{2}\int d^2\ty\tr_Q\Big[ (\tX + \tX_{0})\bdel(\tZ+ \tZ_{0}) - (\tZ+ \tZ_{0})\bdel(\tX + \tX_{0})\Big] \Bigg]
\ee 
(with $\tX_{0,ab}=\inv{N}\frac{u_a\de(\ty-z_a)}{\bdel} \psid_{ka}\psi_{kb},\ \tZ_{0,ba}= \inv{N}\frac{\de(\ty-z_b)}{\bdel}\psid_{bl}\psi_{al}$) where $\tZ_{ab}, \tX_{ab}$ are Grassmann even $Q\times Q$ matrices, parametrizing open string modes on the $D1'$ branes, see Fig. \ref{fig:openstrings}. Inserting this into \eqref{ch_psi}, we can cancel the quartic fermion term, and write 
\be \label{ch_psi_rho}
\begin{split}
    G_Q &= \inv{Z_Q}\int D\tX D\tZ d\psid d\psi \exp\bigg[ -\frac{N}{2}\int d^2\ty\, \tr_Q(\tX\bdel\tZ - \tZ\bdel \tX) - \tX_{ab}(z_b)\psid_{kb}\psi_{ka} \\
    &\hspace{230pt}+ u_a\tZ_{ba}(z_a)\psid_{la}\psi_{lb}+ \sum_a m_a \psid_{ka}\psi_{ka} \bigg] .
\end{split}
\ee 

    \paragraph{Integrating out the fundamental fields:} Now we can integrate out $\psi, \psid$, and write the correlator as a functional integral involving only the open strings on $D1'$ branes\footnote{Going back to the asymmetric form for simplicity.} 
\be \label{ch_rho}
G_Q = \inv{Z_Q} \int D\tX D\tZ \exp\bigg[-N\int d^2\ty\, \tr_Q(\tX(\ty)\bdel\tZ(\ty))\bigg] \det(\mc{K}(u_a, z_b))^N,
\ee 
where $\mc{K}_{ab}(u_b,z_b) = m_a\de_{ab} - \tX_{ab}(z_b) + u_b\tZ_{ab}(z_b)$\footnote{Note that $\tX, \tZ$ are functions of $\ty$, and $\tX(z_a)\equiv\tX(\ty = z_a)$ and $\tZ(z_a)\equiv \tZ(\ty=z_a)$.}. 

We see that the correlation function of $Q$ determinants in the original theory ( $Q$ probe branes in the presence of $N$ coincident branes) can be expressed as a one point function of a heavy ($\Delta \sim \mc{O}(NQ)$) composite operator in the theory on the probe branes. This is our first main result 
\be 
\begin{split}
    &\hspace{150pt}\text{\textbf{Result 1a:}}\\
    &\underbrace{\inv{Z_N} \int DXDY\, \exp\bigg[ -N\int d^2z\, \tr_N(X\bdel Y) \bigg]\prod_{a=1}^Q \det(m_a + \mc{W}(u_a, z_a))}_{\text{V-type field theory with $Q$ probe branes}}\\
    &\hspace{80pt}= \underbrace{ \inv{Z_Q} \int D\tX D\tZ \exp\bigg[-N\int d^2\ty\, \tr_Q(\tX\bdel\tZ)\bigg] \det(\mc{K}(u_a, z_b))^N}_{\text{F-type field theory on the probe branes}}
\end{split}
\ee

\section{The Dual $\b\ga$ System and Backreacted Geometry}\label{coinc}
\begin{figure}
    \centering
    \includegraphics[width=0.7\linewidth]{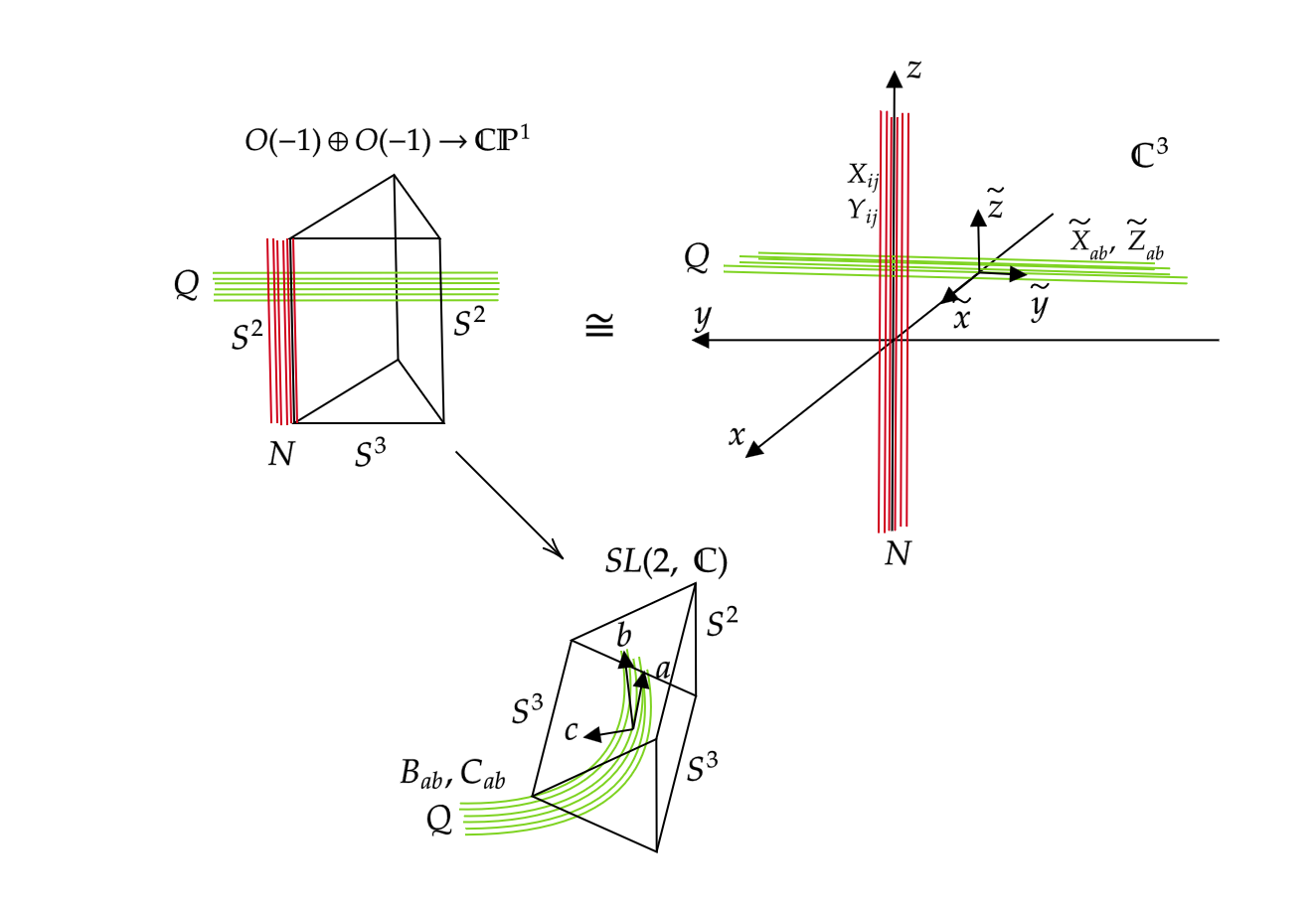}
    \caption{Coincident branes and backreaction. (Left) In addition to the $N$ branes (in red), we have taken the $Q$ probe branes (in green) to be coincident. (Right) Same setup in the $\mb{C}^3$ patch. (Bottom) The backreaction from the $N$ branes changes the complex structure to the deformed conifold, where the probe branes follow a curve. }
    \label{fig:probe2}
\end{figure}

We saw in the last section that the integrating in and out procedure leads us to a field theory in terms of a new coordinate $\ty$ and fields $\tX, \tZ$, but their physical meaning is not clear from the expression \eqref{ch_rho}. We will now show that in a certain limit, they can be identified with the coordinate on the probe brane and its transverse fluctuations. To see this, we make the following simplifications: take $m_a = m, u_a = 0, \ \forall a$, and then the limit $z_a\to 0, \ \forall a$. From the point of view of the original geometry, this means we are inserting a stack of coincident $D1'$  branes along some curve $x + m = 0, z=0$ in the resolved conifold \footnote{In order to be able to do the integrating in and out smoothly, we need to take the limits $z_a\to 0, u_a \to 0$ carefully.}, see Fig. \ref{fig:probe2} \footnote{The general scenario (i.e. $u\neq 0, z\neq 0$) can be easily obtained by translation and rotation of the $x,y,z$ coordinates.}.  Then \eqref{ch_rho} takes a simpler form 

\begin{comment}
which can be written, after the rescaling $(\tX, \tZ) \to \sqrt{\frac{N}{Q}}(\tX, \tZ)$
\be 
G_Q = \inv{Z'(\tX, \tZ)} \int D[\tX, \tZ] \exp\Big[ -Q\int d^2\ty \tr_Q(\tX\bdel \tZ) \Big] \det\Big(m - \sqrt{\frac{N}{Q}}(\tX_{kl}(\ty_0)- u\tZ_{kl}(\ty_0))\Big)^N
\ee   
\end{comment}
\be \label{coinc_f_2}
\begin{split}
    G_Q &= \inv{Z_Q} \int D\tilde{X} D\tZ \exp\bigg[-N\int d^2\ty\, \tr_Q(\tilde{X}(\ty)\bdel\tZ(\ty))\bigg] \det(m - \tilde{X}(0))^N\\
    &= \inv{Z_Q} \int D\tilde{X} D\tZ \exp\bigg[-N\int d^2\ty\, \tr_Q(\tilde{X}(\ty)\bdel\tZ) + N\tr_Q \log(m - \tilde{X}(0))\bigg] 
\end{split}
\ee 
We can immediately identify from the action \eqref{coinc_f_2} that $\ty$ represents the coordinate along the stack of $Q$ probe branes, and $\tilde{X}, \tZ$ valued in the adjoint of $U(Q)$, correspond to its fluctuations in the transverse directions $\tilde{x}, \tilde{z}$ marked in Fig. \ref{fig:probe2}. The quadratic piece is the expected $\b\ga$ action, as expected from the full open string field theory discussed in \ref{sec:OSFT}. Furthermore the determinants in \eqref{coinc_f_2} represent the original $N$ branes, inserted at  $\tilde{x}= m, \ty = 0$\footnote{Notice that, from the point of view of the $N$ original branes, the probe branes are at $x +m=0$, but from the point of view of the probe branes, the original branes are at $\tilde{x} -m=0$, which explains the sign difference in the two determinants, see Fig. \ref{fig:probe2}.}. We thus establish a more symmetric version of the first result whose interpretation is somewhat clearer
\be 
\begin{split}
    &\hspace{160pt}\text{\textbf{Result 1b:}}\\
    &\underbrace{\inv{Z_N} \int DXDY\, \exp\bigg[ -N\int d^2z\, \tr_N(X\bdel Y) \bigg] \det(m + X(0))^Q}_{\text{V-type action with $Q$ dets(branes)}}\\
    &\hspace{50pt}=\underbrace{\inv{Z_Q} \int D{\tX} D\tZ \exp\bigg[-N\int d^2\ty\, \tr_Q(\tilde{X}\bdel\tZ)\bigg] \det(m - {\tX}(0))^N}_{\text{F-type action with $N$ dets(branes)}}
\end{split}
\ee 
In the coincident limit of the probe branes, we therefore recover the gauged $\b\ga$ description of the open strings, but on the probe branes, with the original $N\times N$ open strings integrated out into a determinant source. 

\subsection*{Field Theory in the Backreacted Geometry}
We see that \eqref{coinc_f_2} is still an action formulated in the undeformed background, i.e. the $N$ branes still appear in the action as determinants. But we will show in this section that we can take the backreaction into account and recast \eqref{coinc_f_2} as the appropriate $\b\ga$ system in the deformed geometry. 

As explained in \cite{CostelloGaiottoTH}, the effect of the backreaction of the $N$ original branes wrapping the holomorphic cycle $\mc{C}$(described by $x = y= 0$ in our case) can be obtained by solving the Kodaira-Spencer (BCOV)\cite{Bershadsky:1993cx, Bershadsky:1993ta} equation with a delta function source (see also \cite{Vafa:2000wi, Costello:2015xsa}): 
\be 
\bdel \a + \half[\a, \a] + N\de_{\mc{C}}=0,
\ee 
whose solution is given in terms of the Bochner-Martinelli kernel
\be \label{beltrami_rho}
\a= \b_{BM}\bdel_z = \frac{N}{4\pi^2}\frac{\bx d\by - \by d\bx}{(|x|^2 + |y|^2)^2}\del_z.
\ee 
\begin{comment}

For transparency, let's also write down the solution in the $\rho, \s, \ty$ coordinate system, where $\mc{C}= \{\tilde{\rho} = \rho + u\s = m, \ty = \ty_0\}$
\be \label{beltrami_rho}
\a = \frac{N}{4\pi^2} \frac{(\bar{\tilde{\rho}} -m)d\bar{\s} - \bar{\s}d\bar{\tilde{\rho}}}{(|\tilde{\rho} -m|^2 + |\s|^2)}\del_{\s}
\ee 
\end{comment}
This unique solution changes the complex structure to the deformed conifold, or equivalently $SL(2,\mb{C})$. Therefore, if the backreaction of the $N$ branes are taken into account, then the theory on the probe branes should be described by a gauged $\b\ga$ system in $SL(2,\mb{C})$. If we represent the $SL(2,\mb{C})$ with coordinates $(a, b, c, d)\in \mb{C}^4$ (with $ad - bc=1$), then we can follow \eqref{holo_CS_bg} and write the gauge-fixed action (ignoring the ghosts) as 
\be \label{def_con_bg}
S = N\int d^2a \inv{a}\tr_Q(B\bdel_a C) 
\ee 
(in the patch $a\neq 0$), where we have assumed that $a$ is the coordinate along the branes and $B, C$ represent the transverse fluctuations. The $\inv{a}$ appears due to the holomorphic 3-form $\Omega$ \eqref{def_con_omega}. 

We will now show that \eqref{coinc_f_2} is equivalent to \eqref{def_con_bg} under a particular identification of coordinates and fields. For the sake of simplicity of expressions, let's simplify \eqref{coinc_f_2} further by taking $m=0$ and start with the action 
\be 
\begin{split}
    &S=N\int d^2\ty\ \tr_Q\Big(\tX(\ty)\bdel_{\ty}\tZ(\ty)\Big) - N \ \tr_Q(\log\tX(0))\\
    &\xrightarrow[\ty\to\pi^{-1}\ty ]{\text{rescaling}} \ S'=\frac{N}{\pi}\Big[ \int d^2\ty\ \tr_Q\Big(\tX(\ty)\bdel_{\ty}\tZ(\ty)\Big) -  \pi \tr_Q(\log\tX(0))\Big]
\end{split}
\ee 
and do the following manipulations
\be \label{curved_bg_man}
\begin{split}
    \pi S'&=N\int d^2\ty\ \Big[\tr_Q(\tX\bdel_{\ty}\tZ) - \pi \ \de^2(\ty) \ \tr_Q(\log\tX(\ty))\Big]= N\int d^2\ty\ \Big[\tr_Q\Big(\tX\bdel_{\ty}\tZ\Big) - \ \bdel_{\ty}(\inv{\ty}) \ \tr_Q(\log\tX(\ty))\Big]\\
    &\xrightarrow[\text{by parts)}]{\text{(integrating}} N\int d^2\ty\ \tr_Q\Big(\tX\bdel_{\ty}\tZ\Big) + N\int \frac{d^2 \ty}{\ty} \  \underbrace{\bdel_{\ty}\tr(\log\tX(\ty))}_{=\tr\tX^{-1}\bdel_{\ty}\tX = -\tr\tX\bdel_{\ty}\tX^{-1}}\\
    =&N\int d^2\ty\ \tr_Q\Big(\tX\bdel_{\ty}\tZ\Big) - N\int \frac{d^2 \ty}{\ty} \tr_Q\Big(\tX\bdel_{\ty}\tX^{-1}\Big)\xrightarrow{rearranging} N\int \frac{d^2\ty}{\ty}\tr_Q\Big(\tX(\ty)\bdel_{\ty}\Big(\ty\tZ(\ty) -\inv{\tX(\ty)}\Big)\Big).
\end{split}
\ee 
The last expression of \eqref{curved_bg_man} is exactly the expected form of a $\b\ga$ system \eqref{def_con_bg} in the deformed conifold under the identifications 
\be \label{coord_match_1}
a = \ty, \ B(a) = \tX(\ty), \ C(a) = \ty\tZ(\ty) - \inv{\tX(\ty)} \hspace{30pt}
\ee 

So we see that the expression \eqref{coinc_f_2} is equivalent to the dynamics of the branes in the backreacted geometry (See Fig. \ref{fig:probe2}). A similar expression was also obtained in \cite{Sharma:2025ntb} by considering the interaction between the probe branes and the closed string background, using the open-closed coupling of holomorphic Chern-Simons theory to Kodaira-Spencer theory \cite{Hofman:2002cw} \footnote{We thank Atul Sharma for several clarifying discussions on this point.}.
\be \label{eq:res2}
\begin{split}
    \text{\textbf{Result 2:}}\hspace{20pt} \underbrace{\pi N\int d^2\ty\ \tr_Q(\tX\bdel_{\ty}\tZ) - N\ \tr(\log\tX(0))}_{\text{probe brane action in resolved conifold}} \ = \underbrace{N\int \frac{d^2a}{a}\tr_Q(B\bdel_a C)}_{\text{probe brane action in deformed conifold}}
\end{split}
\ee 

We have thus observed that the change of variables \eqref{coord_match_1} implements the geometric transition exactly at the level of the action\footnote{We have not checked whether the relation between the actions \eqref{eq:res2} translates to the equivalence of the two quantum theories, one would need to study the Jacobian factors coming from \eqref{coord_match_1} carefully to analyze that.}.

\subsubsection*{Relation to Complex Structure Deformation}
%\ds{Can modify/remove. But I feel the validity of the change of coordinates should be shown, and its relation to the kodeira spencer solution. }
Before moving on, we should examine the validity of this change of variables. The relation \eqref{coord_match_1} can be completed and we can describe the deformed conifold in patches using the resolved conifold ($\cong \mb{C}^3$ locally) coordinates (for the general case $m\neq0$) 
\begin{gather} \label{coord_match_2}
    1.\ \{\text{patch } V_1: \tilde{x}-m\neq 0\}\Longrightarrow 
        \begin{bmatrix}
            a & b\\
            c & d
        \end{bmatrix}=\begin{bmatrix}
            \ty & \tilde{x} -m\\
            \ty\tilde{z} - \inv{\tilde{x}-m} & \tilde{z}(\tilde{x}-m)
        \end{bmatrix} \in SL(2,\mb{C})
\end{gather}
\begin{gather}
    2.\ \{\text{patch } V_2:\ty\neq 0\} \Longrightarrow 
        \begin{bmatrix}
            a & b\\
            c & d
        \end{bmatrix}=\begin{bmatrix}
            \ty & \tilde{x} -m\\
            \ty\tilde{z}  & \tilde{z}(\tilde{x}-m) + \inv{\ty}
        \end{bmatrix}\in SL(2,\mb{C})\ \ \ \ \ \ \
\end{gather}
We can verify that they are well defined everywhere except $(\tilde{x} -m=0, \ty=0)$, i.e. at the locus of the $N$ original branes, reflecting the fact that the blown up $\cp$ no longer sits inside the deformed conifold.  On the overlap of the two patches, the transition function is nontrivial, and is encoded in 
\be 
\Big(\frac{c(V_2)}{\ty} - \frac{c(V_1)}{\ty}\Big) = \Big(\frac{d(V_2)}{\tilde{x}-m} - \frac{d(V_1)}{\tilde{x}-m}\Big) = \inv{\ty(\tilde{x}-m)}.
\ee 
This transition function captures the change in complex structure dictated by \eqref{beltrami_rho}\footnote{Or more precisely the solution in the $\tilde{x}, \tilde{y}, \tilde{z}$ coordinate system, i.e.
\be 
\a = \frac{1}{4\pi^2} \frac{(\bar{\tilde{x}} -m)d\bar{\ty} - \bar{\ty}d\bar{\tilde{x}}}{(|\tilde{x} -m|^2 + |\ty|^2)}\del_{\tilde{z}}.
\ee 
}. So this is not some ad hoc change of fields, rather it directly reflects the change in complex structure due to the backreaction of the $N$ branes.

\section{Relation to Matrix Models}\label{matrix}
\subsection{The F-type $\rho$ Matrix Model}
A dual matrix model theory on the probe branes is known in the literature \cite{Budzik:2021fyh}\footnote{The authors of \cite{Lopez-Raven:2024vop} also realized the description of the shape of the giant graviton branes, and their Chan-Paton bundles. It would be interesting to see the relation of their work to ours.} We will show here how their matrix model arises as the on-shell reduction of the field theory  \eqref{ch_psi_rho} and \eqref{ch_rho}. 

Let's start with \eqref{ch_rho} with the symmetric kinetic term
\be 
G_Q = \inv{Z_Q} \int D\tX D\tZ \exp\Big[ - \frac{N}{2}\int d^2\ty \ \tr_Q(\tX\bdel\tZ - \tZ\bdel\tX)\Big] \underbrace{\det(m_a\de_{ab}- \tX_{ab}(z_b) +u_b\tZ_{ab}(z_b))^N}_{\det(\mc{K})^N}
\ee 
The equations of motion of the fields can be written as 
\be \label{ch_rho_eom}
\begin{split}
    \bdel\tZ^{cl}_{ab} &=  -\de(\ty- z_a)\mc{K}^{-1}_{ab} \implies \tZ^{cl}_{ab}(z) = -\inv{\pi(\ty-z_a)}\mc{K}^{-1}_{ab}\\
    \bdel \tX^{cl}_{ab} &= -u_i\de(\ty-z_a)\mc{K}^{-1}_{ab} \implies \tX^{cl}_{ab}(z) = -\frac{u_a}{\pi(\ty-z_a)}\mc{K}^{-1}_{ab},
\end{split}
\ee
where we are only keeping track of singular pieces, and ignoring holomorphic constants of integration.

Let's compute the on-shell action next
\be 
\begin{split}
    S_{on-shell} &= \frac{N}{2}\int d^2\ty\ \tr_Q(\tX^{cl}\bdel\tZ^{cl} - \tZ^{cl}\bdel\tX^{cl}) -N\tr_Q\log(\mc{K})\\
    &= \frac{N}{2}\int d^2\ty \Big(\sum_{a\neq b}\frac{u_{ab}\de(z-z_b)}{\pi(z-z_a)}\mc{K}^{-1}_{ab}\mc{K}^{-1}_{ba}\Big)-N\tr_Q\log(m_b\de_{ab} -\frac{u_a}{\pi z_{ab}}\mc{K}^{-1}_{ab} +\frac{u_b}{\pi z_{ab}}\mc{K}^{-1}_{ab})\\
    &= -\frac{N}{2}\sum_{a\neq b} \frac{u_{ab}}{\pi z_{ab}}\mc{K}^{-1}_{ab}\mc{K}^{-1}_{ba} - N\tr_Q \log(m_a\de_{ab} - \frac{u_{ab}}{\pi z_{ab}}\mc{K}^{-1}_{ab}) 
\end{split}
\ee 
If we denote $\rho_{ab} :=-\frac{u_{ab}}{\pi z_{ab}}\mc{K}^{-1}_{ab} $, then the on-shell action can be written as 
\be \label{ch_dual_matrix}
\begin{split}
    &\hspace{75pt}\text{\textbf{Result 3:}}\\
    S_{on-shell} &=  -\frac{N}{2}\sum_{a\neq b}^{Q}\frac{\pi z_{ab}}{u_{ab}}\rho_{ab}\rho_{ba} - N\tr_Q \log(m_a\de_{ab} + \rho_{ab}).
\end{split}
\ee 
or, 
\be \label{ch_dual_matrix_2}
G_Q \sim \inv{Z[\rho]} \int [d\rho] e^{-S[\rho]}\sim \inv{Z[\rho]} \int [d\rho]\exp\Big( \frac{N\pi}{2}\sum_{a\neq b} \frac{z_{ab}}{u_{ab}}\rho_{ab}\rho_{ba} + N\tr_Q \log(m_a\de_{ab} + \rho_{ab})\Big)
\ee 
This is precisely the dual matrix model action derived in \cite{Budzik:2021fyh}. The underlying relation \eqref{ch_rho_eom} between the matrix variable $\rho$ and the fields $\tX, \tZ$ can be simply written as 
\be \label{ch_mm_eom}
\tX_{ab}(\ty)\vert_{on-shell}= u_a\tZ_{ab}(\ty)\vert_{on-shell} = \frac{u_a}{\ty-z_a}\frac{z_{ab}}{u_{ab}}\rho_{ab}.
\ee
This proves our third main result. As conjectured in \cite{CostelloGaiottoTH}, the backreaction from the $N$ original branes changes the complex structure of the Calabi-Yau to the deformed conifold, equivalently $SL(2,\mb{C})\sim AdS_3 \times S^3$. One remarkable result of \cite{Budzik:2021fyh} is that the saddle point solution of \eqref{ch_dual_matrix} can be represented as a giant graviton brane solution in the deformed conifold, thus supporting the conjecture. For each saddle solution of $\rho$, they showed that the following commuting matrices 
\be \label{sl2c_saddle}
\begin{split}
    B(a) &= a\mu - (\rho + m)\\
    C(a) &= a\zeta +(\rho + m)^{-1}\\
    D(a) &= a\mu\zeta - \zeta(\rho+ m) + \mu(\rho + m)^{-1}
\end{split}
\ee 
satisfy $aD(a)- B(a)C(a)=1$, hence their eigenvalues describe curves in $SL(2,\mb{C})$. This is the classical shape of the branes. Now using the relation \eqref{ch_mm_eom}, we see that the classical shape of the giant graviton brane arises from the on-shell value, or the $vev$ of the fields $\tX, \tZ$. We are not sure how the $\inv{N}$ corrections will change the description of the giant graviton branes. 

\begin{comment}
    This should be related to the transverse fluctuation of the brane in the deformed geometry. Let's look at the expansion 
\be 
\begin{split}
    S' &= S - S_{on-shell}\\
    &= \int d^2z(\de P_{ij}\bdel\de\tZ_{ji}) - \tr \log'(\de_{ik} - (M + \chi)^{-1}_{ij}(\de P_{jk}(z_k) - u_k\de\tZ_{jk}(z_k)))\\
    &= \int d^2z \tr_Q(\de P\bdel\de\tZ) - \sum_{n=2}^{\infty} \frac{(-1)^n}{n}\tr\Big((M+\chi)^{-1}(\de P - u\de\tZ)\Big)^n
\end{split}
\ee 
where $\log'$ essentially means the power series in the second line starting from $n=2$, since the linear term cancels due to expanding around an extrema. As we can see, the action is very nonlocally sourced, which is expected as we started with probe branes inserted at different loci in $\mb{C}^3$. To get an action analogous to the $\beta-\gamma$ system on $SL(2, \mb{C})$, we need to take the branes to be nearly coincident. 
\end{comment}

\subsection{The V-type Matrix Model}
The relation between the dual field theory expression \eqref{ch_rho} and the dual matrix model expression \eqref{ch_dual_matrix} makes us wonder if the original expression of the correlator of determinant operators \eqref{ch_Z} can be reduced to a matrix integral as well. We will show that this is indeed the case. 

Let's start with \eqref{ch_Z},
\be 
G_Q =  \inv{Z_N} \int DXDY \, \exp \bigg[ -\frac{N}{2} \int d^2z\ \tr_N( X\bdel Y - Y\bdel X)  \bigg]\prod_{a=1}^Q \det(m_a + \mc{W}(u_a, z_a)).
\ee 
Equations of motion of the adjoints are given by  
\be 
\bdel Y^{cl}_{ji}(z) = \inv{N}\sum_a \de(z-z_a)(m_a + \mc{W}(u_a,z))^{-1}_{ji} \, \implies Y^{cl}_{ji}(z) = \inv{\pi N}\sum_a \frac{(m_a + \mc{W}(u_a, z_a))^{-1}_{ji}}{z-z_a}.
\ee 
Let's call $(m_a + \mc{W}(u_a, z_a))^{-1}_{ji} =N\pi M_{a, ji} $ , then we can write 
\be \label{Z_eom1}
Y^{cl}_{ji}(z) = \sum_a\frac{M_{a,ji}}{(z-z_a)}.
\ee 
Similarly, we have 
\be \label{Z_eom2}
X^{cl}_{ji}(z) = -\inv{N\pi}\sum_a\frac{u_a}{z-z_a}(m_a + \mc{W}(u_a, z_a))^{-1}_{ji} = -\sum_a\frac{u_a M_{k,ji}}{(z-z_a)}.
\ee 
Now let's look at the on-shell action
\be 
\begin{split}
    S_{on-shell} &= \frac{N}{2}\int d^2z \,\tr_N(X^{cl}\bdel Y^{cl} - Y^{cl} \bdel X^{cl}) - \sum_a \tr_N \log\Big(m_a + X^{cl}(z_a) + u_a Y^{cl}(z_a)\Big)  \\
    &= \frac{N\pi}{2}\sum_{a\neq b} \frac{u_{ab}}{z_{ab}} M_{a, ij}M_{b, ji} -\sum_a \tr_N\log\Big(m_a\de_{ij} - \sum_{b\neq a}\frac{u_{ab}}{z_{ab}}M_{b, ji} \Big) 
\end{split}
\ee 
(Note that there are some diverging terms in the on-shell action, but they precisely cancel out). Written in terms of the on-shell action and variables $M_{a, ij}$, we can reduce the correlator \eqref{ch_Z} to that of a multi-matrix model, namely
\be \label{ch_og_matrix}
G_Q \sim \inv{Z[\{M_a\}]}\int [DM_a] \exp\Big[ -\frac{N\pi}{2}\sum_{a\neq b}\frac{u_{ab}}{z_{ab}} \tr_N(M_a M_b) \Big] \prod_{a=1}^{Q}\det\Big(m_a - \sum_{b\neq a}\frac{u_{ab}}{z_{ab}}M_{b}\Big)
\ee 
The duality between the matrix models of the form \eqref{ch_og_matrix} and \eqref{ch_dual_matrix_2} were explored in detail in \cite{gopakumar2023deriving, Gopakumar:2024jfq}. We see now that they can arise as an on-shell reduction of gauge theories describing open strings on different kinds of D-branes. 

\begin{comment}
An alternate way to show the same equivalence is to take \eqref{ch_psi}, and integrate in the following matrix chain integral 
\be 
1 = \inv{{Z[\{Z_k\}]}}\int D[\psid, \psi] D[Z_k]\exp\Bigg[-\frac{N\pi}{2}\sum_{k\neq l}\frac{u_{kl}}{z_{kl}} (Z_{k,ba} + \inv{N\pi}\psid_{kb}\psi_{ka} )(Z_{l,ab} +\inv{N\pi}\psid_{la}\psi_{lb} )\Bigg]
\ee 
so that the quartic fermion terms cancel, and we can integrate out the fermions from the remaining quadratic action to get precisely \eqref{ch_og_matrix}. 
\end{comment}

\section{Discussion \& Outlook}\label{conc}
\subsubsection*{Summary and Main Takeaways}
In this paper, we have taken the first step towards understanding the open-closed-open triality of \cite{gopakumar2023deriving} beyond matrix models, namely in the twisted holography setting of \cite{CostelloGaiottoTH}. By studying a system of two stacks of branes in the resolved conifold, we have found two equivalent descriptions of the open strings stretched between one set of branes, while treating the other stack as a potential. Both descriptions are full-fledged field theories, and are really the open string field theory description \`a la \cite{WittenCSasOSFT}. Furthermore we have shown that our dual field theory action, with the effective potential, correctly captures the backreaction of the $N$ original branes, that deforms the complex structure from resolved to deformed conifold. This can be thought of as using an open string probe to detect the backreacted closed string geometry. Finally we have also shown that the $\rho$-matrix model arises as an effective theory, with the on-shell action of the field theory matching precisely the action of the $\rho$-model. This disentangles the field- versus matrix-integral descriptions from the V- and F-type nature of the dualities discussed in \cite{gopakumar2023deriving, Gopakumar:2024jfq}.

\subsubsection*{Future Directions}

\begin{itemize}
    \item \textbf{The full F-type theory:} As mentioned in the introduction, we are far from deriving the full F-type picture of twisted holography as envisioned in \cite{gopakumar2011simplest, freefieldsIII}. A possible next step towards that goal can be to consider higher-dimensional branes wrapping complex surfaces in the resolved conifold, which should be some sort of generalized determinants in the OSFT/gauge theory language, perhaps along the lines of \cite{Ishtiaque:2018str}. These higher-dimensional branes were recently discussed in connection to the giant graviton expansion in \cite{Budzik:2025vzr}. It would be interesting to explore this in the future. 
    \item \textbf{Generalize to Free $\mathcal{N}=4$ SYM:} We can carry out a very similar analysis as above for free SYM, as in \cite{komatsuOCO}. In Appendix \ref{sym}, we derive a simple field theory on the probe branes as well, not just the matrix model discussed in the literature so far. It is the theory of six free scalars with an effective potential generated by the integrating in-out procedure of open-closed-open triality. They should correspond to the transverse fluctuations of the $Q$ $D3'$-branes, and by a similar logic, the potential term should somehow encode the AdS-geometry. The situation will likely be complicated by the stringy bulk effects at zero coupling. We leave this for future work.
    \item \textbf{Graph Duality of Feynman Diagrams \& connection to Two-Matrix Model:} The duality of field theories discussed here are very similar to the web of matrix model dualities discussed in \cite{Gopakumar:2024jfq}. There, by inserting two families of determinants at one of two positions in $\mathcal{N}=4$ SYM, we found multiple interrelated descriptions in terms of different open strings. It also gave us an idea of how the underlying A-model string theory looks like. It would be interesting to carry out something similar here, and use it as a guiding principle to flesh out an A-model worldsheet description of twisted holography. 
\end{itemize}

\section*{Acknowledgements}
It is our pleasure to thank Kasia Budzik, Frank Coronado, Matthias Gaberdiel, Rajesh Gopakumar, Rishabh Kaushik, Shota Komatsu, Ji Hoon Lee, Wei Li, Nathan McStay and Atul Sharma for helpful discussions and feedback on the draft. EAM acknowledges support by a SwissMAP Research fellowship and the ITP of ETH Zurich. DS acknowledges support by U.S. Department of Energy grant DE-SC0007870, and Harvard University. 

\pagebreak

\appendix

\section{Free $\mc{N}=4$ Super Yang-Mills Theory}\label{sym}
In this appendix, we carry out a similar analysis as the main text, in the free $\mc{N}=4$ super Yang-Mills theory. We will find some interesting new results similar to the main text, but leave their physical interpretation for future work. 

We are interested in $\half $ BPS operators in $\mc{N}=4$ SYM. These are created from traces and products of traces of a null linear combination $Y \cdot \Phi = Y_I \Phi^I, Y\cdot Y =0$, so the operators look like 
\be
\prod_{i=1}^n \tr_N(Y\cdot \Phi(x))^{l_i}
\ee 
We can also consider heavier BPS operators, given by (sub-)determinants
\be 
\mc{D}(m, Y)(x) = \det(m + Y\cdot \Phi(x))
\ee 
These operators correspond to sphere giant gravitons in the dual $AdS_5 \times S^5$ geometry \cite{McGreevy:2000cw, Hashimoto:2000zp, Balasubramanian:2001nh, Corley:2001zk}. The quantities we will be interested in are correlators of these determinant insertions in the free limit of the theory, 
\be \label{sym_det_corr_1}
\langle \prod_{k}\mc{D}(m_k, Y_k)(x_k)\rangle_{g_{YM}=0}
\ee 
Note that while two and three point functions of $\half$ BPS operators are expected to be protected by supersymmetry \cite{Lee:1998bxa, Eden:2000qp, Eden:2000gg}, hence we can obtain the actual correlation function from the free theory computation. But a general n-point function is not expected to be protected, so a free field computation won't match with the actual result. Rather, \eqref{sym_det_corr_1} will be the answer in free super YM theory.

\subsection*{Determinant Correlators}
The determinant correlator in the free theory limit can be written as 
\be \label{YM_phi}
G_m = \frac{1}{Z_{\Phi}}\int [D\Phi^{I}] \left(\prod_k \mathcal{D}_k \right) \,\,\exp\left[-\frac{1}{g_{\rm YM}^2}\int d^4 x\,\,{\rm tr}_N\left( \del_{\mu}\Phi^{I}\del^{\mu}\Phi_I\right)\right]\comma
\ee 
Next we express the determinants in terms of Grassmann odd bifundamentals \cite{komatsuOCO}
\be
\mathcal{D}_k =\det (m_k + Y_k\cdot \Phi)=\int d\psid_{ka}d\psi_{ka} \exp \left[\psid_{ka} (m_k \de_{ab} + Y_k\cdot \Phi_{ab}(x_k))\psi_{kb}\right]\comma
\ee
Next we can follow the same idea as before and integrate out the adjoint fields $\Phi^I$, obtaining the following expression
\beq\label{YM_psi}
G_m=\int d\psi_kd\psi_k\,  \exp \left[-\frac{g^2}{N}\sum_{k\neq l} \frac{(Y_k\cdot Y_l)}{|x_{kl}|^2}(\psid_{kb}\psi_{lb}\psid_{la}\psi_{ka}) + m_k\psid_{ka}\psi_{ka} \right]\comma
\eeq
where we have rewritten things in terms of the 't Hooft coupling
\be 
g^2\equiv\frac{\lambda}{16\pi^2}=\frac{g_{\rm YM}^2 N}{16\pi^2}\period
\ee

\subsection*{Theory on the Probe Branes}
We can ask, just like the previous case, whether the correlation function can be expressed in terms of the field theory defined on the probe branes. And the way to obtain the theory is again similar to the above analysis. We integrate in six $Q\times Q$ fields $\rho^I, \ I=1, ..,6$ and insert the following quantity in \eqref{YM_psi} 
\be \label{YM_rho_psi}
\begin{split}
    1 &= \inv{Z_\rho} \int [D\rho] \exp \bigg[  -\frac{N}{g^2}\sum_{k\neq l} \int d^4x (\rho^I_{kl}(x) - \frac{g^2}{2\pi N}\frac{(\psid_{kb} Y^I_k \psi_{lb})}{|x-x_k|^2})\Box (\rho^I_{lk}(x) - \frac{g^2}{2\pi N}\frac{(\psid_{la} Y^I_l \psi_{ka})}{|x-x_l|^2})\bigg]\\
    &= \inv{Z_\rho} \int [D\rho] \exp \bigg[ -\frac{N}{g^2}\int d^4x {\rm tr}_Q(\rho^I\Box\rho^I) -4\pi \rho^I_{kl}(x_l)(\psid_{la} Y^I_l \psi_{ka})+\frac{g^2}{N} \sum_{k \neq l}  \frac{Y_k\cdot Y_l}{|x_{kl}|^2} (\psid_{kb} \psi_{lb} \psid_{la} \psi_{ka}) \bigg]
\end{split}
\ee
where we have used the identity 
\beq\label{eq:dividentity}
\Box\frac{1}{|x|^2}=-4\pi^2 \delta^4 (x)
\eeq
We can see the fermion quartic term in \eqref{YM_rho_psi} cancels the one in \eqref{YM_psi}, and hence we obtain
\be \label{YM_rho}
\begin{split}
    G_m &= \inv{Z_\rho} \int [D\rho] d\psid_k d\psi_k \exp \bigg[ -\frac{N}{g^2}\int d^4x\,  {\rm tr}_Q(\rho^I\Box\rho^I) -4\pi \rho^I_{kl}(x_l)(\psid_{la} Y^I_l \psi_{ka}) + m_k\psid_{ka}\psi_{ka} \bigg]\\
    &= \inv{Z_\rho} \int [D\rho] \exp \bigg[ -\frac{N}{g^2}\int d^4x\,  {\rm tr}_Q(\rho^I\Box\rho^I) \bigg] {\rm det}(m_k\de_{kl}-4\pi\rho_{lk}(x_k)\cdot Y_k)^N
\end{split}
\ee 
So we can see that the determinant correlator \eqref{YM_phi} can be written as a one point function of a heavy nonlocal operator. We suspect, motivated by the discussion in the main text, that \eqref{YM_rho} in the correct coincident limit can be translated to the action of giant graviton branes in $AdS_5\times S^5$. 

\subsubsection*{Dual (F-type) Matrix Model}
The expression \eqref{YM_rho} can be related to the matrix model obtained in \cite{komatsuOCO}, using the on-shell action, just like in the case of twisted holography. Going through similar steps as in the main text, we obtain
\be 
G_m\vert_{on-shell} = \inv{Z[\bar{\rho}]}\int [D\bar{\rho}]\exp\Bigg[\frac{N}{4\pi^2g^2}\tr(\bar{\rho}^2)\Bigg] \det\Big(m_k\de_{kl} + \inv{\pi}\sqrt{\frac{Y_k\cdot Y_l}{|x_{kl}|^2}}\bar{\rho}_{lk}\Big)^N
\ee 
which is exactly the matrix model obtained in \cite{komatsuOCO}. As pointed out to us by S. Komatsu, we can obtain a saddle equation for this matrix model, which looks very similar to \cite{Budzik:2021fyh}
\be \label{eom}
\inv{g^2}[\mc{X}, [\mc{X}, \bar{\rho}]] + [\mc{Y}, [\mc{Y}, \bar{\rho}^{-1}]]\sim 0
\ee 
It would be interesting to see whether \eqref{eom} leads to the shape of giant graviton branes in $AdS_5\times S^5$. 

\subsubsection*{Original (V-type) Matrix Model}
Similarly the original V-type expression \eqref{YM_phi} can be reduced to a multimatrix model through an on-shell analysis, given by the expression
\be 
\begin{split}
    G_m\vert_{on-shell} &=\inv{Z[Z]}\int dZ \exp\Bigg[ \frac{N}{4g^2}\sum_{k\neq l} \Big( \frac{Y_k\cdot Y_l}{|x_{kl}|^2}\Big)\tr_N(Z_k Z_l)\Bigg] \prod_{k=1}^{Q} \det\Big(  m_k\de_{ab} -\sum_{l\neq k}\frac{Y_k\cdot Y_l}{|x_{kl}|^2}Z_{l, ba} \Big)
\end{split}
\ee 
We see the appearance of a $Q$-matrix chain, very similar to \eqref{ch_dual_matrix_2}. We see that the usual open-closed-open string duality of matrix models \cite{gopakumar2023deriving, Gopakumar:2024jfq} can be obtained from a similar duality in the context of free Yang-Mills theories.

\bibliographystyle{utphys}
\bibliography{refs}
\end{document}